\documentclass[structabstract,a4paper]{aa}  
\usepackage{graphicx}
\usepackage{natbib}
\usepackage{lscape}
\usepackage{rotating}
\usepackage{txfonts}

\newcommand{\kms}{km\ s$^{-1}$}

\newcommand{\xmm}{{\sc XMM}\emph{-Newton}}

\newcommand{\zp}{$\zeta$\,Puppis}

\begin{document}
  \title{A detailed X-ray investigation of \zp \thanks{Based on observations collected with XMM-Newton, an ESA Science Mission with instruments and contributions directly funded by ESA Member States and the USA (NASA).}}
   \subtitle{I. The dataset and some preliminary results}

   \author{Ya\"el Naz\'e\inst{1}\fnmsep\thanks{Research Associate FRS-FNRS}
           \and Carlos Arturo Flores\inst{2}
           \and Gregor Rauw \inst{1}
          }

   \institute{GAPHE, D\'epartement AGO, Universit\'e de Li\`ege, All\'ee du 6 Ao\^ut 17, Bat. B5C, B4000-Li\`ege, Belgium\\
              \email{naze@astro.ulg.ac.be}
   \and Departamento de Astronomia, Universidad de Guanajuato, Apdo Postal 144, Guanajuato, GTO, Mexico
             }


 
  \abstract
   {} 
   {\zp, one of the closest and brightest massive stars, was the first early-type object observed by the current generation of X-ray observatories. These data provided some surprising results, confirming partly the theoretical predictions while simultaneously unveiling some problematic mismatches with expectations. In this series of papers, we perform a thorough study of \zp\ in X-rays, using a decade of \xmm\ observations.}
   {\zp\ was observed 18 times by \xmm, totaling 1Ms in exposure. This provides the highest-quality high-resolution X-ray spectrum of a massive star to date, as well as a perfect dataset for studying X-ray variability in an ``archetype'' object. }
   {This first paper reports on the data reduction of this unique dataset and provides a few preliminary results. On the one hand, the analysis of EPIC low-resolution spectra shows the star to have a remarkably stable X-ray emission from one observation to the next. On the other hand, the fitting by a wind model of individual line profiles recorded by RGS confirms the wavelength dependence of the line morphology.}
   {}


   \maketitle
%

\section{Introduction}
With its very early spectral type (O4Infp, \citealt{wal72}) and a distance of only 335\,pc \citep{van07,mai08}, the star Naos, better known as \zp\ (or HD\,66811), is one of the closest and brightest massive stars. It is therefore one of the most studied objects amongst the O-star population. However, despite the intense work, many open questions remain on its nature.

Indeed, \zp\ displays several intriguing properties. First, its visible spectrum shows clear signs of helium overabundance and chemical enrichment by CNO-processed material (e.g. \citealt{pau01}) as well as fast rotation (more than 200\kms\ for $v \sin(i)$, \citealt{pen96,how97}). Second, it is a known runaway (e.g. from Hipparcos data, \citealt{mof98}). These properties have led to speculations on its evolutionary status. On the one hand, the chemical enrichment and fast rotation could result from mass and angular momentum exchange through Roche lobe overflow in a binary. \zp\ could therefore have been the secondary component of such a system, the supernova explosion of its companion having ejected it from its birth place a few millions years ago \citep{van96}. On the other hand, \zp\ {displays a similar Hipparcos parallax} as stars of the Vela R2 association\footnote{ This conclusion was based on the original release of the Hipparcos catalog, hence the use of the `old' distance of 430\,pc in the Schaerer et al. paper, but the parallax similarity remains when using the new reduction of Van Leeuwen (and thus the `new' distance of 335\,pc).} \citep{sch97}, and dynamical interactions within this association could have led to the ejection of the (single) O-star \citep{van96}. In this scenario, the chemical enrichment of \zp\ would be explained by the intense rotational mixing occurring in the fast-rotating main-sequence progenitor \citep{mey00}. In addition, \zp\ displays double-peaked emission lines, suggested to arise in a rotating wind \citep{con74,pet96}, and a compression of the wind in the equatorial plane was detected by \citet{har96}.

Due to its brightness, \zp\ was one of the first massive stars observed with high-resolution in X-rays \citep{kah01,cas01}. At first, its X-ray lines appeared to match expectations as they did show the broad, blueward-skewed profiles expected for the wind embedded shock model \citep{owo01}. However, the devil was in the details. When quantitatively fitting the line profiles, \citet{kra03} found a much lower wind attenuation than expected on the basis of the mass-loss rate determined from optical and UV observations (see also \citealt{osk06}). They also found that the typical optical depths $\tau_*$, used in the wind-shock models, seemed independent of wavelength, which can only be explained by invoking porosity \citep{fel03, osk06}. To improve the fitting of the X-ray line profiles, \citet{leu07} included the effect of resonance scattering: better fits were indeed obtained, without the need of a large reduction in the mass-loss rate, but they also showed that some unexplained discrepancies remain. Re-analyzing the {\it Chandra} data of \zp, \citet{coh10} argue in favor of a reduced mass-loss rate, without the need of any porosity as their new derivation of the optical depths implies an increase with wavelength, as expected from the bound-free absorption opacity of the (cool) wind. Except for \citet{leu07}, all above studies relied on a single 68\,ks {\it Chandra} observation or a 57\,ks \xmm\ exposure taken in 2000.  Both facilities have their advantages: while \xmm\ globally has a higher sensitivity, {\it Chandra} has a lower background, and a higher spectral resolution and sensivity at short wavelengths for its grating spectra.  Today, however, much more data are available (see below).

Considering the uncertainties in the line profile results and the lack of new variability studies, we decided to re-investigate \zp\ using the best dataset available at the present time: 18 \xmm\ exposures, corresponding to an exposure of 1Ms totaling $>$700\,ks of useful time (i.e. an improvement by an order of magnitude compared to most previous studies). This dataset thus provides the most detailed X-ray view of an O star to date. The results that we obtained will be presented in a series of papers. The first one will present the data, their reduction, and a few first results; the second one will focus on the X-ray variations of \zp, using EPIC and RGS data; the last one will present a global analysis of the high-resolution X-ray spectrum, using the merged high-resolution data.

This first paper is organized as follows. The dataset and its reduction are presented in Sect. 2, the spectral fits are presented in Sect. 3, the individual line profile fitting in Sect. 4, and the results are summarized and discussed in Sect. 5.

\section{\xmm\ observations}
In the past decade of \xmm\ observations, the star \zp\ was observed 18 times, mostly for calibration purposes. These datasets are excellent for studying the variability of \zp\ since (1) the scheduled exposure times were often long (up to $\sim$60\,ks) and (2) the observing dates probe weekly, monthly, and yearly timescales. Unfortunately, many observations were affected by soft proton background flares, resulting in total exposure times reduced by about 30\%.  Total net exposure times for EPIC-MOS, EPIC-pn, and RGS amount to 579\,ks, 477\,ks,\, and 751\,ks. A summary of the observations is given in Table \ref{journal}. The successive columns provide the dataset ID (obsID and revolution number); the date at mid-exposure (in the format dd/mm/yy + UT time and JD--2\,450\,000., calculated using the start/end times of the observations listed in the on-line 'observation lokator'); the mode as well as the scheduled, performed, and effective (i.e. after cleaning flares) exposure time for both EPIC-MOS and EPIC-pn instruments; the scheduled, performed, and effective exposure time for RGS. An empty column indicates a discarded or unavailable dataset (see below). Note that the target was placed off-axis in two observations (5.95\arcmin\ off-axis in Rev. 0731 and 1.1\arcmin\ off-axis in Rev. 0903). 

\subsection{EPIC data}

The EPIC data were reduced with SAS v10.0.0 using calibration files available on January 1 2011 and following the recommendations of the \xmm\ team\footnote{SAS threads, see \\ http://xmm.esac.esa.int/sas/current/documentation/threads/}. Different modes (small window, large window, full frame, timing) as well as different filters (thick, medium) and position angles were used for these observations, resulting in a somewhat heterogeneous dataset. To ensure the most homogeneous analysis, hence a meaningful comparison between datasets, two decisions were taken. First, a few observations were discarded: those when the instruments were not ``on'' (aka $CAL\,CLOSED$), those totally affected by flares, those with the source appearing totally or mostly in a CCD gap, those with very short exposure times ($<$10\,ks), those taken in timing mode, and those using a unique combination of mode+filter. This trimming process results in a final dataset composed of 9 observations taken with large window + thick filter and 6 observations taken with small window + thick filter for EPIC-MOS; 10 observations taken with small window + thick filter, 5 observations taken with large window + medium filter mode and 4 observations taken with large window + thick filter for EPIC-pn (see Table \ref{journal}). Second, the extraction regions were chosen to be as constant as possible. A single source region was used for EPIC-MOS, whatever the mode, but two background regions were defined, one for each mode since it was not possible to extract the background on the same CCD chip as the source for the small window mode. For EPIC-pn, a single source region was used for all modes, as well as a single background region for the large window mode but four different background regions were necessary for the small window mode. Table \ref{reg} gives the position and shape of each of these regions.

\subsubsection{Pile-up}
\zp\ is rather bright in X-rays, with EPIC count rates of $\sim$2\,cts\,s$^{-1}$ (for MOS) and $\sim$6.5\,cts\,s$^{-1}$ (for pn with thick filter). These count rates are at the pile-up limit from the XMM Users' handbook for the large window modes (i.e. 1.8 and 6\,cts\,s$^{-1}$ for MOS and pn, respectively) but well below the limits for the small window mode (i.e. 5 and 50\,cts\,s$^{-1}$ for MOS and pn, respectively). Some pile-up may thus affect our EPIC data taken in the large window mode. To see how severe the pile-up is, we performed several checks. First, we inspected the event files: no event with $PATTERN$=26--29 was found. The pile-up is thus moderate. Second, we run the SAS task $epatplot$: some small but significant deviation from the ``no pile-up'' configuration is detected for the large window data, especially those taken with the medium filter. 

To get rid of the pile-up, we could extract the data in an annulus centered on the source. This would require the annulus to be perfectly centered on the source, especially since the PSF is far from being symmetric. Getting the exact position of \zp\ in the datasets is however an impossible task. Indeed, the detection algorithm is also disturbed by pile-up, so that the position found in this way for  \zp\ is not accurate. For example, the pipeline-processed data from Rev. 1620 yields a position some 2.5'' away from the Hipparcos position of \zp\ (this separation is the maximum found in the pipeline-processed data), while a dedicated run of the detection algorithm using only the $PATTERN$=0 events gives a position that is only 1.2'' away. One would immediately think of using nearby X-ray sources associated with well-known stars - sources which are less bright in X-rays (thus unaffected by pile-up) but still bright enough to get an accurate position in each dataset. However, such nearby sources do not exist in the neighbourhood of \zp. A perfect centroiding of annular regions is thus impossible and using always the Hipparcos position for annular regions may alter the source's properties, hence the results of the variability study that we wish to perform.

\clearpage
\begin{sidewaystable}
\caption{Journal of the observations. }
\label{journal}
\centering
\begin{tabular}{lc|cc|lccc|lccc|ccc}
            \hline\hline
ObsID & Rev. & Mid-exp. Date & JD & \multicolumn{4}{c|}{EPIC-MOS1}& \multicolumn{4}{c|}{EPIC-pn}& \multicolumn{3}{c}{RGS1}\\
 & & & & Mode & Sched. & Perf.& Real & Mode & Sched. & Perf. & Real & Sched. & Perf. & Real \\ 
 & & & --2\,450\,000.& & (ks) & (ks)& (ks) & (ks) & (ks) & (ks) & (ks)&  (ks)& &  (ks)\\ 
\hline
0095810301 & 0091& 2000-06-08T09:32:39& 1703.898&          &     &     &     &           &     &     &     & 57.4& 57.4& 36.2\\
0095810401 & 0156& 2000-10-15T06:43:44& 1832.780& LW+thick & 37.7& 37.7& 37.3& LW+medium & 35.7& 35.7& 33.4& 40.6& 40.6& 39.9\\
0157160401 & 0535& 2002-11-10T23:40:41& 2589.487& LW+thick & 42.2& 42.2& 41.7& LW+thick  & 13.0& 13.0& 12.1& 42.4& 42.4& 41.6\\
           &     &                    &         &          &     &     &     & LW+medium & 24.4& 24.4& 22.7&     &     &     \\
0157160501 & 0538& 2002-11-17T07:03:34& 2595.794& LW+thick & 43.4& 41.1& 32.2& LW+thick  & 15.7& 15.7& 14.6& 43.6& 42.5& 29.8\\
           &     &                    &         &          &     &     &     & LW+medium & 23.0& 23.0& 12.2&     &     &     \\
0157160901 & 0542& 2002-11-24T20:26:10& 2603.352& LW+thick & 43.4& 43.4& 42.9& LW+thick  & 14.1& 14.1& 13.2& 43.6& 43.6& 43.0\\
           &     &                    &         &          &     &     &     & LW+medium & 24.6& 24.6& 20.9&     &     &     \\
0157161101 & 0552& 2002-12-15T04:53:31& 2623.704&          &     &     &     & LW+medium & 24.2& 24.0& 11.5& 45.6& 38.9& 26.9\\
0159360101 & 0636& 2003-05-30T19:28:01& 2790.311& LW+thick & 66.8& 62.7& 18.8& SW+thick  & 42.8& 42.7& 24.3& 72.9& 69.2& 56.2\\
0163360201 & 0731& 2003-12-07T02:47:04& 2980.616&          &     &     &     & LW+thick  & 61.2& 52.6& 32.4& 62.9& 53.6& 35.8\\
0159360301 & 0795& 2004-04-12T17:33:58& 3108.232& LW+thick & 63.9& 41.8& 19.0& SW+thick  & 30.2& 30.2& 17.4& 64.1& 61.3& 21.1\\
0159360401 & 0903& 2004-11-14T01:57:57& 3323.582& LW+thick & 21.9& 21.9& 21.6& SW+thick  & 29.8& 29.8& 20.9& 77.0& 63.0& 48.2\\
0159360501 & 0980& 2005-04-16T14:39:28& 3477.111& LW+thick & 29.3& 29.3& 29.0& SW+thick  & 63.8& 63.8& 22.1& 64.2& 64.2& 31.0\\
           &     &                    &         & SW+thick & 34.1& 27.7& 13.4&           &     &     &     &     &     &     \\
0159360701 & 1071& 2005-10-15T04:04:52& 3658.670&          &     &     &     & SW+thick  & 59.6& 22.2& 15.5& 60.0& 30.0& 27.5\\
0159360901 & 1096& 2005-12-04T01:14:14& 3708.552& SW+thick & 59.8& 53.5& 46.2& SW+thick  & 59.6& 53.3& 33.1& 60.0& 53.5& 43.1\\
0159361101 & 1164& 2006-04-17T21:48:48& 3843.409& LW+thick & 58.0& 42.9& 40.1&           &     &     &     & 58.2& 52.9& 40.5\\
0414400101 & 1343& 2007-04-09T22:49:29& 4200.451& SW+thick & 63.7& 63.7& 47.3& SW+thick  & 63.5& 63.5& 34.2& 63.9& 63.9& 48.8\\
0159361301 & 1620& 2008-10-14T01:15:08& 4753.552& SW+thick & 66.2& 61.2& 53.3& SW+thick  & 66.0& 61.0& 38.3& 66.4& 61.5& 54.7\\
0561380101 & 1814& 2009-11-04T06:17:00& 5139.762& SW+thick & 64.1& 64.1& 62.1& SW+thick  & 63.9& 63.9& 44.7& 64.3& 64.3& 60.5\\
0561380201 & 1983& 2010-10-07T23:09:52& 5477.465& SW+thick & 76.7& 76.7& 74.3& SW+thick  & 76.5& 76.5& 53.5& 76.9& 76.9& 65.9\\
\hline
\multicolumn{2}{l|}{Total exposure time} & &    &          &771.2&679.9&579.2&           &791.6&734.0&477.0&1066.7&979.7&750.7\\
\hline	
\end{tabular}
\tablefoot{EPIC-MOS2 and RGS2 have similar, though sometimes not identical, exposures as EPIC-MOS1 and RGS1, respectively. }
\end{sidewaystable}
\clearpage

\begin{table*}
\caption{Regions used for extracting source and background data from EPIC instruments. }
\label{reg}
\centering
\begin{tabular}{lccccr}
            \hline\hline
Inst.+mode & Src/Bkgd & shape & RA$_{\rm center}$ & DEC$_{\rm center}$  & radii \\
& & & (hh:mm:ss)& (dd:mm:ss) & (px)\\
\hline
All    & Src  & circle  & 08:03:35.047 & $-$40:00:11.33 & 850 \\
MOS+LW & Bkgd & circle  & 08:03:40.322 & $-$40:02:18.62 & 600     \\
MOS+SW & Bkgd & circle  & 08:02:57.173 & $-$39:56:40.39 & 1000    \\
pn+LW  & Bkgd & circle  & 08:03:42.928 & $-$39:57:28.61 & 700     \\
pn+SW  & Bkgd1& circle  & 08:03:35.047 & $-$40:00:11.33 & 700     \\
       & Bkgd2& circle  & 08:03:29.219 & $-$40:02:51.12 & 700     \\
       & Bkgd3& circle  & 08:03:48.368 & $-$39:58:58.59 & 700     \\
       & Bkgd4& circle  & 08:03:42.928 & $-$39:57:28.61 & 700     \\
\hline
\end{tabular}
\tablefoot{Here, 1\,px=0.05\arcsec. The source position is from the Hipparcos catalog (cf. Simbad). The background regions Bkgd 2 to 4 were used for Revs. 0636+0795+0980+1343, Revs. 0903+1071+1096, and Revs. 1620+1814+1983, respectively. In all other cases, the background region Bkgd1 was used.}
\end{table*}

On the other hand, there is an alternative way in cases of mild pile-up: using only $PATTERN$=0 events. After such a filtering, we compared the $PATTERN$=0 spectra extracted in a circular region with the spectra extracted in annular regions (hence free of pile-up) using the usual $PATTERN$=0--12 for MOS and $PATTERN$=0--4 for pn. This check was done by fitting simple 2-temperatures models on LW+medium data from Rev. 0156, LW+thick data from Rev. 0731 and SW+thick data from Rev. 1814 for EPIC-pn, and on LW+thick data from Rev. 0156 and SW+thick data from Rev. 1814 for EPIC-MOS. The comparison is excellent for EPIC-MOS data: fluxes and count rates differ by $<$1\% and best-fit spectral parameters are within the errors. The remaining difference can be attributed to the slightly larger noise in the spectra extracted in annuli and from calibration differences. The comparison is less perfect for EPIC-pn, especially for the data taken with the medium filter: the flux differences reaches 6\% in this case, and best-fit spectral parameters are at 2-$\sigma$ from each other. The pile-up thus still has a small influence on the EPIC-pn data taken in the large window mode, and those data should thus be considered with caution.

\subsubsection{The final files}

A final check was made on the data from Rev. 1620, which yields the most discrepant (2.5'') position for \zp, if we trust the pipeline processing. We first derived the position of \zp\ from the $PATTERN$=0 data using the SAS task $edetectchain$, and then extracted the spectra using a circular region centered on that position. We compared these spectra to those extracted on $PATTERN$=0 data using a circular region centered on the Hipparcos position of \zp. Both sets of spectra appear identical in Xspec: the small centroiding errors have thus no impact on the spectra as long as a circular region is used. 

We therefore extracted lightcurves and spectra of \zp\ in a circular region centered on the Hipparcos position of the target. We used only the $PATTERN$=0 event files. While this is not necessary for the small window mode, it ensures a homogeneous data reduction. Xspec v12.6.0 was used to fit spectra, and our own software to analyze EPIC lightcurves.

Note that spectra were grouped using the new SAS task $specgroup$. It enables to reduce the oversampling, which may ``cause problems during spectral fitting because the spectral bins are then not completely independent'' (excerpt from SAS 10.0.0, online documentation). We choose an oversampling factor of 5, ensuring that no spectral bin is narrower than 1/5 of the full width half maximum resolution at the central photon energy of the bin. Note that, while providing more statistically correct data, this process dramatically reduces the number of spectral bins. The data were also grouped to ensure that a minimum signal-to-noise of 3 was reached in each spectral bin of the background-corrected spectra.

\subsection{RGS data}
The RGS datasets were also reduced in a standard way with SAS v10.0.0. Many new, important RGS features were modified in that version (e.g. the spectral binning in wavelength rather than in dispersion angle units). This ensures a better calibration of our datasets. It also solved the calibration problems (wavelength shift and reduced flux) found when using earlier versions of the SAS for the two observations where \zp\ was placed off-axis. Note that the data were extracted using the proposal position of the source, which is the same in all observations but the first two (Revs. 0091 and 0156, shift of 0.0002$^{\circ}$ in both RA and DEC) - this small position shift has no impact on the derived RGS spectra.

When detected, flares were discarded using $rgsfilter$\footnote{Rev. 1071 has an increasingly high background towards the observation's end, but no ``discrete'' flare. The whole observing time was therefore used.}. The tasks {\it rgsspectrum} and {\it rgsrmfgen} then provided unbinned source and background spectra, as well as response matrices for each order (1,2) and each instrument (1,2). A final, combined spectrum was also calculated using all 18 RGS datasets and the task $rgscombine$. The background files and matrix responses were attached to the source spectra using the new SAS task $specgroup$, which we also use to ensure an oversampling factor of maximum 5 (see above).

Fluxed spectra combining both RGS instruments and both orders were obtained using the task $rgsfluxer$. Note that a correction for off-axis angles is applied to ensure that the fluxes are real photon fluxes and not simply recorded count rates (i.e. the arf response matrix is fully taken into account). The spectra of one revolution were sampled to get 1500 spectral bins, while the spectrum combining the 18 datasets was calculated to get 3000 spectral bins. This ensures an oversampling factor of about 3 and 6 for the former and latter cases, respectively. Two caveats should be noted. First, the rmf matrix is not fully taken into account by $rgsfluxer$ and there is no instrumental width correction. This needs to be accounted for when modelling the spectra (see Paper III). Second, there are known small wavelength shifts in RGS spectra, apparently depending on the Sun aspect angle. However, no sign of such an effect is detected in our dataset when we use SAS v10 (they would appear as small spectral variations, and there are none, see Paper II).

\section{EPIC spectra}

With data of such high quality, the error bars on the spectra are very small, and it is therefore very difficult to get a formally acceptable fit. We thus avoided to try to get a perfect fit (i.e. $\chi^2\sim1$), which is actually impossible to get without going into unrealistic, overcomplicated models (e.g. 10 components fits with independent, free abundances) when all instruments agree. Rather, we have tried to get a fit as simple as possible which is at the same time as realistic as possible and as close as possible to the spectral data. 

\citet{zhe07} showed that the high-resolution {\it Chandra} spectrum can be fitted by a shock model where the dominant temperature is 0.1--0.2\,keV (\zp\ was actually the star with the coolest dominant plasma), though some contribution from plasma with 0.3--0.7\,keV was needed to achieve a good fit.  This modelling clearly shows that the plasma in \zp\ is rather cool. We decided to use a simpler formalism, which does not try to reproduce shocks, but simply considers the addition of optically-thin thermal plasma (without any assumption on their origins). In Xspec, the models using a distribution of thermal plasma (e.g. $c6pvmkl$) fail to provide a fit close to the data: we therefore had to fit the data using a sum of individual optically-thin plasmas. As could be expected, one, two or three temperature fits do not provide fits close to the data, and similar conclusions are reached for four-temperature fits with a single absorption (which is unsurprising in view of the $c6pvmkl$ result). Our model of choice is thus a four temperature model with individual absorptions: $tbabs\times \sum vphabs\times vapec$, where the first component represents the interstellar absorption, fixed to 8.9$\times 10^{19}$\,cm$^{-2}$ \citep{dip94}, and the abundances of the absorption and emission components are assumed to be equal.  Adding a fifth thermal components (e.g. at 1\,keV) does not significantly improve the fit, and we thus sticked to the decision of using 4 components. 

\begin{figure}
\includegraphics[height=8cm,angle=270]{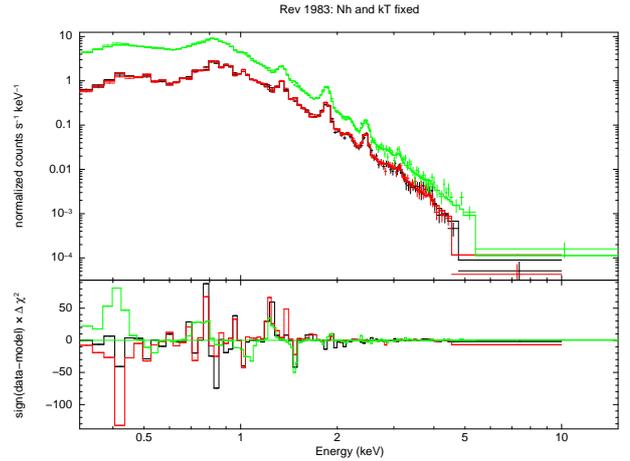}
\caption{The EPIC pn (top, in green) and MOS (bottom, in red and black) data of Rev. 1983 superimposed on the best-fit model.}
\label{epicspec}
\end{figure}

After the best-fit temperatures and absorptions were found, we investigated the impact of using non-solar abundances. We began by keeping the helium abundance to 2 times solar \citep{rep04}, and let the nitrogen and oxygen abundances vary. We perfom such a simultaneous fit to all available EPIC spectra of each revolution. A few conclusions could be drawn from this trial. First, the pn spectra obtained in Large Window mode with the Medium filter (Revs 0156 and 0552, and second pn observation of Revs 535, 538 and 542) are clearly deviant from other data, showing the impact of pile-up. Since all other data (pn or MOS, LW or SW modes with Thick filter) overall agree with one another, we only discarded the pn data taken with Medium filter from further analysis. Second, the absorptions and temperatures of the fits do not vary much:  we therefore fixed them to $kT_i$ of 0.09, 0.27, 0.56 and 2.18\,keV and $N_{{\rm H},i}$ of 0.1, 0.1, 0.71, 0. $\times 10^{22}$\,cm$^{-2}$. Finally, the origins of the high $\chi^2$ can be better pinpointed. On the one hand, MOS and pn data do not always agree, especially at 0.4\,keV (see Fig. \ref{epicspec}). This explains why some fits deviate from the mean behaviour: Revs 0731 and 1071 only provided pn data, while Revs 0156 and 1164 consist only of MOS data. This difference between MOS and pn is probably due to remaining cross-calibration problems, not pile-up since \zp\ is far from the pile-up limit for data taken in SW mode with the Thick filter. On the other hand, all three instruments (MOS1, MOS2, and pn) sometimes display $\delta \chi^2$ of the same sign. This is often the case near strong lines, which points toward two explanations: (1) atomic parameters are imperfect (even for APEC - Fig. \ref{epicspec}  shows two examples: at 1.24\,keV, there is flux in the data but not the model - a line is probably missing; at 1.5keV, there is a line in the model but not the data) and (2) the asymetric line shape of \zp\ influence even EPIC data.  We cannot do much for the former, and the latter will actually be studied in detail in the third paper of this series - it is thus beyond the scope of this contribution.

The next step is to free more abundances, but this could yield erratic and/or problematic and/or not better results. For example, the carbon abundance clearly gets  unrealistic (whether freed last or first). Indeed, low-resolution, broad-band spectra such as those taken by EPIC yield few constraints on the carbon abundance and, as often happens with such data, the fitting procedure favors high values of carbon enrichment, whatever its actual value. From previous studies, it is well known that nitrogen is overabundant in \zp, and carbon may be subsolar (e.g. 0.35 times solar in \citealt{pau01} and 0.6 times solar in \citealt{osk07}). However, reported abundances vary quite a lot in literature (see other examples below) and, following \citet{zhe07}, we decided to keep it to solar as for other non-constrained elements. For other elements, freeing the abundance may not improve the fitting quality or may not yield abundances significantly non-solar: this was the case of magnesium, sulphur, iron and neon when released one after another. The silicon abundance stays close to solar but does improve the $\chi^2$, it was thus allowed to vary together with He, N, and O. 

The best fit results are shown in Table \ref{fit}. The parameter errors are taken from the raw fitting results of Xspec  (i.e. these 1-$\sigma$ errors are ``calculated from the second derivatives of the fit statistic with respect to the model parameters at the best-fit'', and are indicative, see Xspec manual). No error is provided for the fluxes. We could indeed use the relative error on count rates, which would yield relative errors in the 0.1--0.3\% range due to the sole Poisson noise. However, relative flux differences between MOS and pn calibrations amount to about 1\% in the soft energy band (where they are maximum), and the use of other similar models also yield 1\% relative errors. Fluxes should therefore be considered as determined to 1/100, not 1/1000, uncertainties.

Note that the listed parameters should not be over-interpretated: they simply represent a convenient way of well fitting the data, no more no less. The abundances, in particular, are indicative, high-resolution data providing much more stringent constraints (see Paper III). It is for example interesting to note that freeing all abundances at the same time could lead to a better $\chi^2$ but to totally erratic and unrealistic abundances (carbon becoming largely overabundant, nitrogen being solar).  This being kept in mind, it is quite remarkable that our results, which should be considered as indicative only, agree rather well with previous abundance determinations. The helium abundance of \zp\ was found to be 1.2 and 3.4 times solar\footnote{As in Xspec: abundances are in number, relative to hydrogen, and relative to solar.} by \citet{pau01} and \citet{osk06}, respectively, and we found an average value of $\sim$2 (though this abundance is only weakly constrained in EPIC spectra). The nitrogen abundance was determined to be 1.7, 6, and 8 times solar by \citet{zhe07}, \citet{osk06}, and \citet{pau01}, respectively, and we found again an average value of $\sim$4. The oxygen abundance is the least constrained of all, with values of 0.75, 1.6, and even 0.16 times solar reported by \citet{pau01}, \citet{osk06}, and \citet{zhe07}, respectively. The latter agrees well with our value, but it is formally unconstrained since its error was 0.23. Our silicon determination also agrees well with the value found by \citet{zhe07}. 

A few general conclusions can be drawn. The dominant components are those with temperatures of 0.09, 0.27\,keV, and 0.56\,keV (providing 20\%, 43\%, and 35\% of the total flux, respectively),  corroborating the conclusion found by \citet{zhe07}, with a different formalism, that cool plasma dominates in \zp. The flux and spectral parameters do not vary much, the largest variations are obtained when only one instrument is available (e.g. Rev. 0156, where the pn data are discarded because of the piled-up associated with the Medium filter). Fluxes (and their associated dispersions) in the total (0.3--4.\,keV), soft (0.3--0.6\,keV), medium (0.6--1.2\,keV), hard (1.2--4.\,keV), and Bergh\"ofer's (0.9--2.\,keV) bands\footnote{For more details on the choice of energy bands, see Paper II.} are 15.3$\pm$0.4, 4.15$\pm$0.14, 8.07$\pm$0.26, 3.06$\pm$0.09, 5.67$\pm$0.17 $\times 10^{-12}$\,erg\,cm$^{-2}$\,s$^{-1}$, respectively. Dispersions typically amount to 3\%, slightly larger than the typical 1\% error, and a shallow decreasing trend is detected, with a decrease of only $\sim$5\% in flux since Rev. 0156. Such a decreasing trend is reminiscent of aging detector sensitivity problems, and we cannot exclude this possibility on the sole basis of the \xmm\ dataset. 

\begin{figure}
\includegraphics[width=9cm]{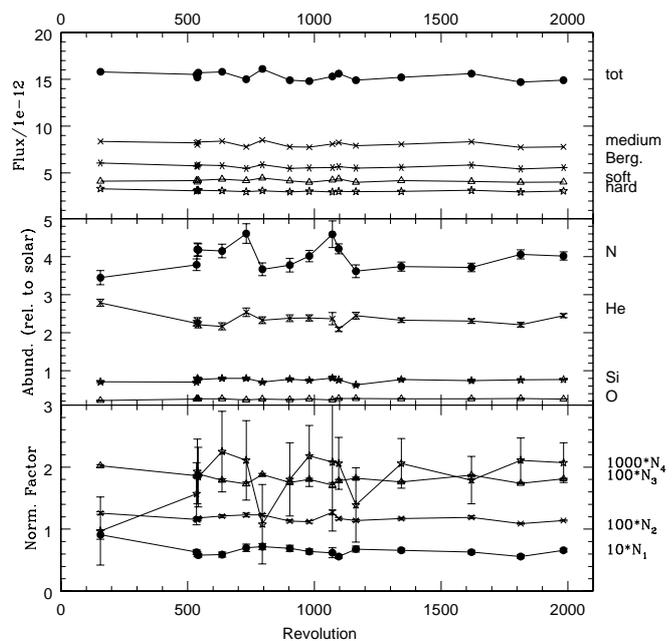}
\caption{Spectral parameters as a function of Revolution number.}
\label{epicmodel}
\end{figure}

\section{Wind profiles}
 \citet{coh10} reported small variations of the line profiles with wavelength,
due to the energy-dependent opacity of the cool wind. However, the {\it Chandra}
observation that they used was relatively short, hence subject to a much higher noise
on the spectrum than for our data, and it is also rather insensitive to
wavelengths $>$20\AA, where the effect is expected to be the largest. 
We therefore re-investigate here the issue using the combined RGS spectrum, 
which has a better signal-to-noise ratio and extends beyond 20\AA. 

\begin{figure*}
\includegraphics[width=8cm]{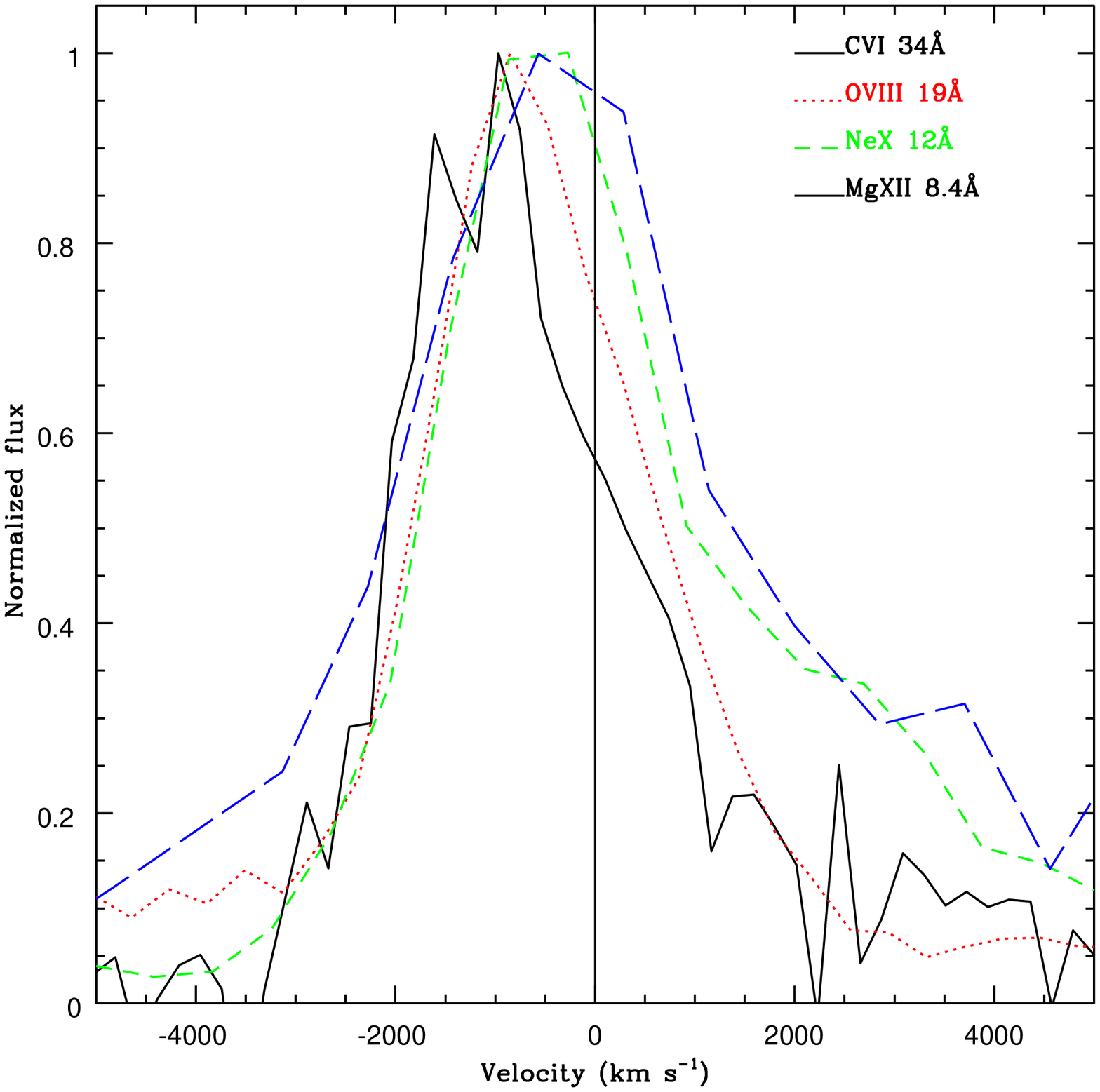}
\includegraphics[width=8cm]{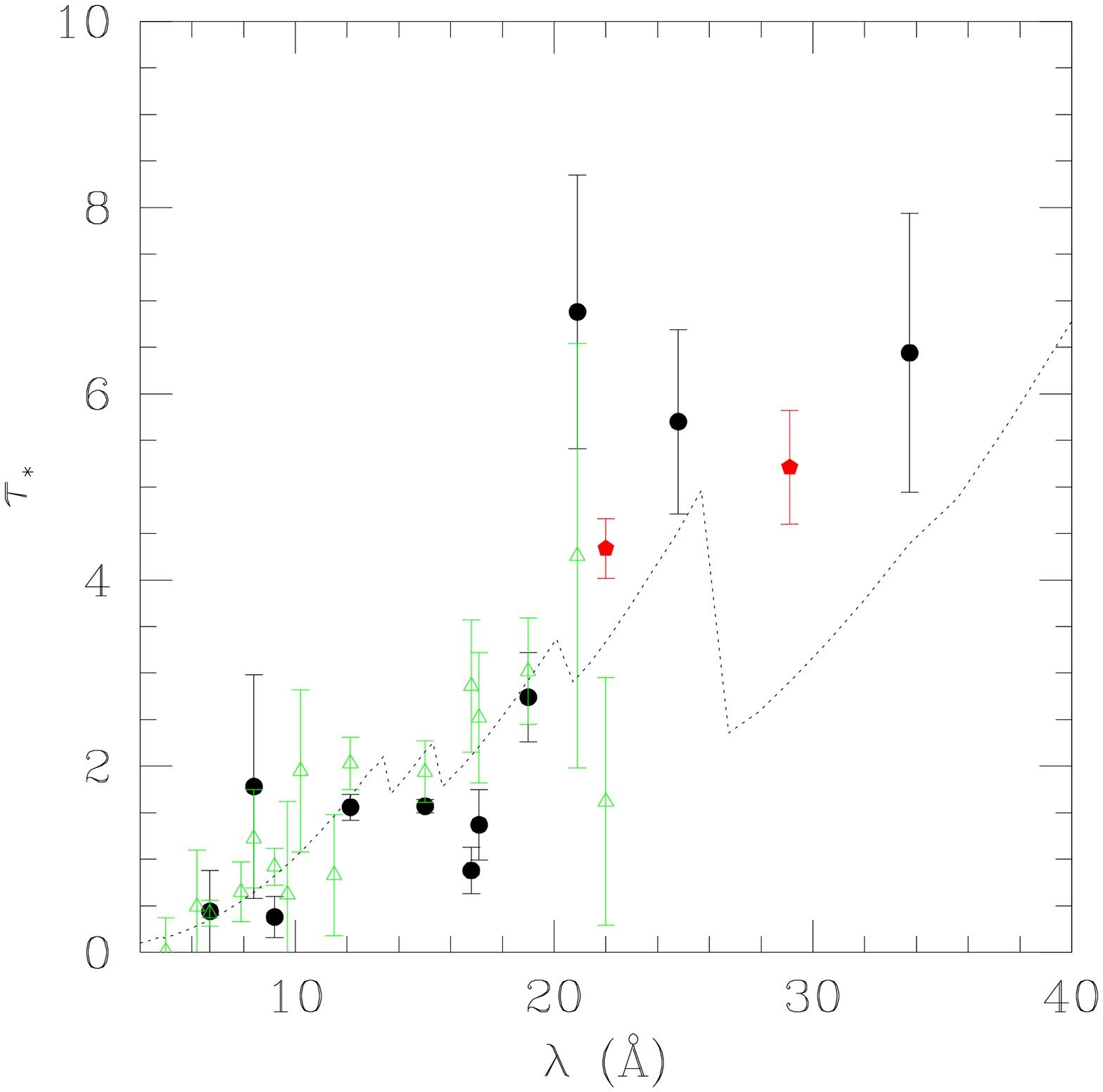}
\caption{Left: Line profiles in velocity space, of the observed Lyman $\alpha$ lines. Right: Variations of the mean optical depth $\tau_*$ with wavelength. Results from this work are shown with filled circles and hexagons - the latter for fits including resonance scattering, while  \citet{coh10} results are displayed with empty triangles. The Cohen et al. theoretical absorption is shown by a dotted line.}
\label{veloc}
\end{figure*}

The left panel of Fig. \ref{veloc} shows the observed Lyman $\alpha$ lines, in velocity space.
For the figure, the lines were approximately continuum-subtracted and normalized 
to have a peak amplitude unity, to highlight their differences. Neighbouring 
lines can be seen for some of these Lyman $\alpha$ lines as bumps in the
blue or red wings. Note that we do not show the N\,{\sc vii} Ly$\alpha$
line as it is blended with a N\,{\sc vi} line.
The variations with wavelength are obvious, as the peak velocity clearly 
appears less blueshifted for the short-wavelength lines. The comparison 
of width and skewness is more difficult by eye, as the RGS resolution
broadens the short-wavelength lines in velocity space, blurring the trends.

To quantify the wavelength variations, we fit the lines with the same 
models\footnote{Wind profile models for Xspec are available on
http://heasarc.nasa.gov/xanadu/xspec/models/windprof.html} as \citet{coh10},
to ensure homogeneity. Results are provided in Table \ref{windprof}: 
the first two columns identify the considered line, the next three columns
define the line shape (characteristic continuum optical depth $\tau_*$,
radius $R_*$ for the onset of the X-ray emission, and the strength of the
line), and the last two column provides details of the line ratios in
the He-like fir triplets. 

Several things must be noted. First, in two cases (the
He-like triplets of N and O), resonance scattering was needed to achieve
a good fit. Second, the Lyman\,$\alpha$ line of nitrogen is blended with
a line from N\,{\sc vi}. We fit these two lines together, assuming that
the line profile parameters are identical: keeping them independent
yields unrealistic results ($\tau_*\sim0$) for the weak N\,{\sc vi} line,
and the lines are so blended that little independent information is
available, explaining the apparently strange results for the weakest line. The 
achieved fit of the nitrogen blend is far from perfect, however ($\chi^2\sim2$). Third,
the iron line at 15\AA\ is not very well fitted ($\chi^2\sim2$), despite
our efforts. It seems that line blends (there are numerous Fe\,{\sc xviii}
lines in the neighbourhood) affect the profile: though these lines
are weak, the very low noise of our data reveals their impact, which 
was not obvious in the {\it Chandra} data. Finally,
the fitting was done using a single terminal velocity for the wind 
(2250\kms, \citealt{pul06}), a $\beta$=1 exponent for the velocity
law, and a power law of zero slope to represent the continuum.

\clearpage
  \begin{sidewaystable}[htb]
  \tiny
  \centering
  \caption{Best-fit parameters.}
  \label{fit}
  \begin{tabular}{l | c c c c | c c c c | c | c c c c c}
  \hline
Rev & $norm_1$ & $norm_2$ & $norm_3$ & $norm_4$ & He & N & O & Si & $\chi^2$ (dof) & \multicolumn{5}{|c}{Observed Flux }\\
& cm$^{-5}$ & 10$^{-2}$\,cm$^{-5}$ & 10$^{-2}$\,cm$^{-5}$ & 10$^{-4}$\,cm$^{-5}$ & & & & & & 0.3-4. & 0.3-0.6 & 0.6-1.2 & 1.2-4.  & 0.9-2.\\
 & & & & & & & & & & \multicolumn{5}{|c}{(10$^{-12}$\,erg\,cm$^{-2}$\,s$^{-1}$)}\\
  \hline
0156 & 0.091$\pm$0.007 & 1.26$\pm$0.02 &  2.02$\pm$0.02 & 0.97$\pm$0.55 & 2.79$\pm$0.09 & 3.45$\pm$0.19 & 0.219$\pm$0.008 & 0.71$\pm$0.02 & 5.48 (163) & 15.8 & 4.12 & 8.35 & 3.31 & 6.06\\
0535 & 0.063$\pm$0.004 & 1.16$\pm$0.02 &  1.86$\pm$0.02 & 1.57$\pm$0.50 & 2.25$\pm$0.07 & 3.79$\pm$0.15 & 0.256$\pm$0.007 & 0.70$\pm$0.02 & 5.79 (266) & 15.5 & 4.16 & 8.22 & 3.07 & 5.75\\
0538 & 0.060$\pm$0.004 & 1.15$\pm$0.02 &  1.85$\pm$0.02 & 1.93$\pm$0.52 & 2.32$\pm$0.08 & 4.19$\pm$0.17 & 0.256$\pm$0.007 & 0.80$\pm$0.02 & 5.17 (268) & 15.2 & 4.02 & 7.98 & 3.13 & 5.71\\
0542 & 0.058$\pm$0.003 & 1.18$\pm$0.02 &  1.89$\pm$0.02 & 1.84$\pm$0.48 & 2.21$\pm$0.08 & 4.18$\pm$0.16 & 0.262$\pm$0.007 & 0.77$\pm$0.02 & 5.76 (276) & 15.7 & 4.20 & 8.31 & 3.16 & 5.85\\
0636 & 0.059$\pm$0.004 & 1.21$\pm$0.02 &  1.79$\pm$0.02 & 2.25$\pm$0.65 & 2.17$\pm$0.09 & 4.15$\pm$0.18 & 0.263$\pm$0.008 & 0.80$\pm$0.02 & 3.37 (257) & 15.8 & 4.34 & 8.38 & 3.09 & 5.77\\
0731 & 0.070$\pm$0.006 & 1.23$\pm$0.03 &  1.73$\pm$0.03 & 2.11$\pm$0.64 & 2.54$\pm$0.11 & 4.61$\pm$0.26 & 0.231$\pm$0.008 & 0.80$\pm$0.03 & 2.92 (111) & 15.0 & 4.19 & 7.78 & 2.98 & 5.48\\
0795 & 0.072$\pm$0.005 & 1.23$\pm$0.02 &  1.88$\pm$0.02 & 1.08$\pm$0.64 & 2.33$\pm$0.09 & 3.67$\pm$0.17 & 0.258$\pm$0.008 & 0.70$\pm$0.02 & 2.78 (249) & 16.1 & 4.45 & 8.51 & 3.09 & 5.86\\
0903 & 0.069$\pm$0.005 & 1.13$\pm$0.02 &  1.75$\pm$0.02 & 1.80$\pm$0.59 & 2.38$\pm$0.09 & 3.78$\pm$0.18 & 0.241$\pm$0.008 & 0.78$\pm$0.02 & 3.06 (259) & 14.9 & 4.13 & 7.78 & 2.97 & 5.49\\
0980 & 0.064$\pm$0.004 & 1.12$\pm$0.02 &  1.80$\pm$0.02 & 2.18$\pm$0.49 & 2.39$\pm$0.08 & 4.02$\pm$0.15 & 0.259$\pm$0.007 & 0.75$\pm$0.02 & 3.33 (414) & 14.8 & 4.00 & 7.74 & 3.04 & 5.53\\
1071 & 0.062$\pm$0.008 & 1.27$\pm$0.04 &  1.71$\pm$0.04 & 2.08$\pm$1.11 & 2.37$\pm$0.16 & 4.59$\pm$0.35 & 0.234$\pm$0.012 & 0.81$\pm$0.04 & 2.02 (96)  & 15.3 & 4.24 & 8.09 & 2.96 & 5.57\\
1096 & 0.056$\pm$0.003 & 1.17$\pm$0.02 &  1.78$\pm$0.02 & 2.06$\pm$0.42 & 2.09$\pm$0.06 & 4.21$\pm$0.13 & 0.273$\pm$0.006 & 0.76$\pm$0.02 & 6.16 (285) & 15.6 & 4.35 & 8.24 & 3.02 & 5.66\\
1164 & 0.068$\pm$0.005 & 1.14$\pm$0.02 &  1.82$\pm$0.02 & 1.39$\pm$0.60 & 2.45$\pm$0.09 & 3.62$\pm$0.17 & 0.274$\pm$0.009 & 0.63$\pm$0.02 & 5.25 (156) & 14.9 & 4.00 & 7.90 & 3.00 & 5.52\\
1343 & 0.066$\pm$0.003 & 1.17$\pm$0.02 &  1.76$\pm$0.01 & 2.06$\pm$0.40 & 2.33$\pm$0.06 & 3.74$\pm$0.12 & 0.265$\pm$0.006 & 0.77$\pm$0.02 & 5.36 (289) & 15.2 & 4.16 & 8.05 & 3.01 & 5.59\\
1620 & 0.063$\pm$0.003 & 1.19$\pm$0.02 &  1.87$\pm$0.01 & 1.79$\pm$0.38 & 2.31$\pm$0.06 & 3.72$\pm$0.11 & 0.264$\pm$0.005 & 0.74$\pm$0.02 & 6.97 (291) & 15.6 & 4.09 & 8.32 & 3.14 & 5.83\\
1814 & 0.056$\pm$0.003 & 1.09$\pm$0.01 &  1.74$\pm$0.01 & 2.11$\pm$0.36 & 2.21$\pm$0.06 & 4.06$\pm$0.12 & 0.269$\pm$0.005 & 0.76$\pm$0.01 & 6.77 (300) & 14.7 & 3.98 & 7.73 & 2.95 & 5.43\\
1983 & 0.066$\pm$0.003 & 1.14$\pm$0.01 &  1.81$\pm$0.01 & 2.07$\pm$0.32 & 2.45$\pm$0.05 & 4.02$\pm$0.11 & 0.257$\pm$0.005 & 0.77$\pm$0.01 & 7.90 (306) & 14.9 & 4.01 & 7.79 & 3.06 & 5.56\\
  \hline
  \end{tabular}
\tablefoot{The fitted model has the form $tbabs\times \sum vphabs\times vapec$, where the interstellar absorption was fixed to 8.9$\times 10^{19}$\,cm$^{-2}$, and the additional absorbing columns and temperatures are fixed to $N_{\rm H,1,2,3,4}$=0.10, 0.10, 0.71, 0.$\times 10^{22}$\,cm$^{-2}$ and $kT_{1,2,3,4}$=0.09, 0.27, 0.56, 2.18\,keV, respectively. Abundances are by number relative to hydrogen and relative to solar.}
  \end{sidewaystable}
\clearpage

\begin{table*}
\caption{Parameters of the wind profiles.}
\label{windprof}
\centering
\begin{tabular}{lcccccc}
            \hline\hline
Ion & $\lambda$ & $\tau_*$ & $R_0/R_*$ & $norm$ & {\cal G}=(f+i)/r & $P=\phi_*/\phi_c$ \\
 & (\AA) &  & & (10$^{-4}$\,ph\,cm$^{-2}$\,s$^{-1}$) &  & (in 10$^4$) \\
\hline
\vspace*{-0.15cm}\\
Si\,{\sc xiii}-tr        & 6.7  & $0.4_{0.0}^{0.8}$ & $1.0_{1.0}^{1.2}$ & $1.15_{1.12}^{1.18}$ & $0.68_{0.55}^{0.79}$ & $0.0002_{0.0001}^{0.0005}$ \\
\vspace*{-0.15cm}\\
Mg\,{\sc xii}-Ly$\alpha$ & 8.4  & $1.8_{0.9}^{3.0}$ & $1.2_{1.0}^{1.5}$ & $0.30_{0.27}^{0.32}$ & & \\
\vspace*{-0.15cm}\\
Mg\,{\sc xi}-tr          & 9.2  & $0.4_{0.2}^{0.6}$ & $1.4_{1.3}^{1.5}$ & $2.05_{2.02}^{2.07}$ & $0.73_{0.67}^{0.78}$ & $0.004_{0.003}^{0.005}$ \\
\vspace*{-0.15cm}\\
Ne\,{\sc x}-Ly$\alpha$   & 12.1 & $1.6_{1.4}^{1.7}$ & $1.5_{1.4}^{1.6}$ & $3.03_{3.00}^{3.06}$ & & \\
\vspace*{-0.15cm}\\
Fe\,{\sc xvii}           & 15.0 & $1.6_{1.5}^{1.6}$ & $1.7_{1.6}^{1.7}$ & $6.36_{6.28}^{6.45}$ & & \\
\vspace*{-0.15cm}\\
Fe\,{\sc xvii}           & 16.8 & $0.9_{0.7}^{1.1}$ & $1.5_{1.5}^{1.6}$ & $2.69_{2.66}^{2.72}$ & & \\
\vspace*{-0.15cm}\\
Fe\,{\sc xvii}           & 17.1 & $1.4_{1.1}^{1.7}$ & $1.5_{1.4}^{1.6}$ & $4.08_{4.06}^{4.12}$ & & \\
\vspace*{-0.15cm}\\
O\,{\sc viii}-Ly$\alpha$ & 19.0 & $2.7_{2.3}^{2.9}$ & $1.0_{1.0}^{1.3}$ & $4.62_{4.53}^{4.72}$ & & \\
\vspace*{-0.15cm}\\
N\,{\sc vii}-Ly$\beta$   & 20.9 & $6.9_{5.6}^{8.3}$ & $1.3_{1.0}^{1.9}$ & $1.80_{1.75}^{1.85}$ & & \\
\vspace*{-0.15cm}\\
O\,{\sc vii}-tr          & 21.7 & $4.3_{4.0}^{4.6}$ & $1.3_{1.0}^{1.6}$ & $7.31_{7.23}^{7.39}$ & $1.06_{1.02}^{1.07}$ & $3.1_{2.4}^{3.8}$ \\
\vspace*{-0.15cm}\\
N\,{\sc vii}-Ly$\alpha$  & 24.8 & $5.7_{4.7}^{6.7}$ & $2.1_{1.1}^{3.2}$ & $6.16_{6.48}^{5.84}$ & & \\
\vspace*{-0.15cm}\\
N\,{\sc vi}-tr           & 29.1 & $5.2_{4.6}^{5.7}$ & $2.7_{2.5}^{2.8}$ & $18.7_{18.6}^{18.9}$ & $1.18_{1.15}^{1.19}$ & $7.6_{5.9}^{9.3}$ \\
\vspace*{-0.15cm}\\
C\,{\sc vi}-Ly$\alpha$   & 33.7 & $6.4_{5.4}^{7.9}$ & $1.1_{1.0}^{1.7}$ & $1.24_{1.20}^{1.29}$ & & \\
\vspace*{-0.15cm}\\
\hline
\end{tabular}
\tablefoot{tr refers to He-like triplets, and Ly to Lyman lines. Shifts and line profiles are identical for all lines of the doublets (Lyman $\alpha$, Lyman $\beta$, Fe\,{\sc xvii} doublet at 17.1\AA) and triplets; the line ratio in doublets is fixed to the ratio of maximum emissivities. For the N\,{\sc vi} triplet, resonance scattering parameters are $\tau_*^0>56$, $\beta_{sob}<0.28$; for the O\,{\sc vii} triplet, resonance scattering parameters are $\tau_*^0>18$, $\beta_{sob}=1.4_{0.9}^{2.0}$. }
\end{table*}

As it is always compatible with zero and as it yields no improvement 
of the fits, no global line profile shift was applied, except for the iron 
lines near 17\AA\ where the improvement is significant with 
a shift of only --15$\pm$2\,\kms.  Wavelength shifts between instruments/orders
were envisaged, but they yielded erratic results generally without improvement of 
the fit, sometimes large values (incompatible with wavelength calibration reports) 
for example for the iron line at 15\AA, values compatible with zero within 3 sigma, and values for one instrument/order compatible with those of another instrument/order within 3 sigma. Cross-correlation also suggests null shifts between exposures 
and between different instrument/order combinations of the same exposure. Due to the broadness of its lines, \zp\ is indeed not the best source to find such shifts: to 
do so, the \xmm\ calibration team uses sources with unresolved X-ray lines. 

As shown in the right panel of Fig. \ref{veloc}, the optical depth varies with wavelength.
Our fitting thus confirms the \citet{coh10} preliminary results by 
extending them to longer wavelengths and by decreasing
the noise on most lines (except the bluest ones).   The wind of \zp\
therefore is unlikely to be composed of clumps which are fully opaque at all wavelengths, as had been suggested for some porosity models. Indeed, optically thick clumps should produce a grey opacity that would solely be determined by the geometry of the clumps.
In principle, these optical depth variations
may be used to constrain the mass-loss rate, for a given star+wind 
model. This was attempted in \citet{coh10}, and they found that a reduced
mass-loss rate with non-solar abundances provided the best-fit results.
 For comparison purposes, the same theoretical opacity is shown in Fig. \ref{veloc} with a dotted 
line: below 20\AA, it indeed provides a good fit, but for larger wavelengths,
the agreement is less good because of the nitrogen edge at 26\AA\ - an edge 
which appears very strong as the nitrogen abundance is enhanced for \zp.
 This discrepancy may be solved by adapting the value of the abundances 
and mass-loss rate of \zp, but this is beyond the scope of this paper.
Clearly, additional modelling is needed to reproduce the full behaviour 
of \zp\ in X-rays  (and this will be done in Paper III).

In contrast, the onset radius appears remarkably stable and
confined in the 1--1.5\,$R_*$ range, which is quite usual in massive stars 
\citep[and references therein]{gud09}. The sole exceptions are the 
Nitrogen triplet and Ly$\alpha$ lines, but it would not be surprising that, 
as these lines have emissivities that peak at rather low temperatures, it simply 
indicates a formation further out in the wind.

\section{Conclusion}
In the past decade, about 1Ms of data were obtained by \xmm\ on \zp. Of these, about 30\% is strongly affected by flares, reducing the useful exposures to 579ks for EPIC-MOS, 477ks for EPIC-pn, and 751ks for RGS. A variety of modes was used, the most reliable being the SW+Thick Filter mode; the use of the medium filter yielded piled-up data, while the data taken in LW+Thick Filter mode do not appear significantly different from those obtained with SW+Thick Filter mode. Attention was paid to this problem, notably by choosing similar extraction regions for all datasets and using only single events. 

Broad-band EPIC data taken with the Thick filter were analyzed using absorbed optically-thin thermal emissions. Four temperatures are needed to reproduce in a reasonable way the data. Note that the fits are not formally acceptable, but that (1) instruments do not always agree with one another and (2) the reduced noise amplifies the limitations due to the imperfect atomic parameters and standard line profiles. Nevertheless, the EPIC spectra appear remarkably stable over the decade of observations, with only 3\% dispersions around the average fluxes. A detailed variability study, based on lightcurves, will be presented in Paper II.

 The combined high-resolution RGS spectrum confirms that the X-ray line profiles vary with wavelength. Fitting individual line profiles using a wind model yields similar onset radius for the X-ray emission, but wind continuum opacities depending on wavelength. This is simply due to the fact that the cool absorbing clumps in the wind are not fully optically thick at all wavelengths, though further modelling is needed in order to adequately reproduce the opacity variations. 

\begin{acknowledgements}
YN acknowledges L. Mahy for his help with CMFGEN, and the \xmm\ helpdesk for interesting discussions about the data. YN and GR acknowledge support from the Fonds National de la Recherche Scientifique (Belgium), the Communaut\'e Fran\c caise de Belgique, the PRODEX XMM and Integral contracts, and the `Action de Recherche Concert\'ee' (CFWB-Acad\'emie Wallonie Europe). CAF acknowledges support from the FNRS-Conacyt agreement, as well as from the `Programa de becas para la formac{\' i}on de j\'ovenes investigatores de la DAIP-GTO'. ADS and CDS were used for preparing this document.
\end{acknowledgements}


\begin{thebibliography}{}


\bibitem[Berghoefer et al.(1996)]{ber96} Berghoefer, T.~W., Baade, D., Schmitt, J.~H.~M.~M., Kudritzki, R.-P., Puls, J., Hillier, D.~J., \& Pauldrach, A.~W.~A.\ 1996, \aap, 306, 899 

\bibitem[Cassinelli et al.(2001)]{cas01} Cassinelli, J.~P., Miller, N.~A., Waldron, W.~L., MacFarlane, J.~J., \& Cohen, D.~H.\ 2001, \apjl, 554, L55 

\bibitem[Cohen et al.(2010)]{coh10} Cohen, D.~H., Leutenegger, M.~A., Wollman, E.~E., Zsarg{\'o}, J., Hillier, D.~J., Townsend, R.~H.~D., \& Owocki, S.~P.\ 2010, \mnras, 405, 2391 


\bibitem[Conti \& Leep(1974)]{con74} Conti, P.~S., \& Leep, E.~M.\ 1974, \apj, 193, 113 


\bibitem[Diplas \& Savage(1994)]{dip94} Diplas, A., \& Savage, B.~D.\ 1994, \apjs, 93, 211 


\bibitem[Feldmeier et al.(2003)]{fel03} Feldmeier, A., Oskinova, L., \& Hamann, W.-R.\ 2003, \aap, 403, 217 



\bibitem[G{\"u}del \& Naz{\'e}(2009)]{gud09} G{\"u}del, M., \& Naz{\'e}, Y.\ 2009, \aapr, 17, 309 

\bibitem[Harries \& Howarth(1996)]{har96} Harries, T.~J., \& Howarth, I.~D.\ 1996, \aap, 310, 533 



\bibitem[Howarth et al.(1997)]{how97} Howarth, I.~D., Siebert, K.~W., Hussain, G.~A.~J., \& Prinja, R.~K.\ 1997, \mnras, 284, 265 

\bibitem[Kahn et al.(2001)]{kah01} Kahn, S.~M., Leutenegger, M.~A., Cottam, J., Rauw, G., Vreux, J.-M., den Boggende, A.~J.~F., Mewe, R., G\"udel, M.\ 2001, \aap, 365, L312 

\bibitem[Kramer et al.(2003)]{kra03} Kramer, R.~H., Cohen, D.~H., \& Owocki, S.~P.\ 2003, \apj, 592, 532 

\bibitem[Leutenegger et al.(2007)]{leu07} Leutenegger, M.~A., Owocki, S.~P., Kahn, S.~M., \& Paerels, F.~B.~S.\ 2007, \apj, 659, 642 

\bibitem[Ma{\'{\i}}z Apell{\'a}niz et al.(2008)]{mai08} Ma{\'{\i}}z Apell{\'a}niz, J., Alfaro, E.~J., \& Sota, A.\ 2008, poster presented at IAU Symposium 250, arXiv:0804.2553 

\bibitem[Meynet \& Maeder(2000)]{mey00} Meynet, G., \& Maeder, A.\ 2000, \aap, 361, 101 


\bibitem[Moffat et al.(1998)]{mof98} Moffat, A.~F.~J., et al.\ 1998, \aap, 331, 949 

\bibitem[Oskinova et al.(2006)]{osk06} Oskinova, L.~M., Feldmeier, A., \& Hamann, W.-R.\ 2006, \mnras, 372, 313 

\bibitem[Oskinova et al.(2007)]{osk07} Oskinova, L.~M., Hamann, W.-R., \& Feldmeier, A.\ 2007, \aap, 476, 1331 

\bibitem[Owocki \& Cohen(2001)]{owo01} Owocki, S.~P., \& Cohen, D.~H.\ 2001, \apj, 559, 1108 

\bibitem[Pauldrach et al.(2001)]{pau01} Pauldrach, A.~W.~A., Hoffmann, T.~L., \& Lennon, M.\ 2001, \aap, 375, 161 

\bibitem[Penny(1996)]{pen96} Penny, L.~R.\ 1996, \apj, 463, 737 

\bibitem[Petrenz \& Puls(1996)]{pet96} Petrenz, P., \& Puls, J.\ 1996, \aap, 312, 195 

\bibitem[Puls et al.(2006)]{pul06} Puls, J., Markova, N., Scuderi, S., et al.\ 2006, \aap, 454, 625 



\bibitem[Repolust et al.(2004)]{rep04} Repolust, T., Puls, J., \& Herrero, A.\ 2004, \aap, 415, 349 

\bibitem[Schaerer et al.(1997)]{sch97} Schaerer, D., Schmutz, W., \& Grenon, M.\ 1997, \apjl, 484, L153 



\bibitem[van Leeuwen(2007)]{van07} van Leeuwen, F.\ 2007, \aap, 474, 653 

\bibitem[van Rensbergen et al.(1996)]{van96} van Rensbergen, W., Vanbeveren, D., \& De Loore, C.\ 1996, \aap, 305, 825 

\bibitem[Walborn(1972)]{wal72} Walborn, N.~R.\ 1972, \aj, 77, 312 


\bibitem[\protect\citeauthoryear{Zhekov \& Palla}{2007}]{zhe07} Zhekov S.A., \& Palla, F. 2007, MNRAS, 382, 1124

\end{thebibliography}
\end{document}